\documentclass[lettersize,journal]{IEEEtran}
\usepackage{cite}

\usepackage{amsmath,amssymb,amsfonts,amsthm}
\usepackage{algorithmic}
\usepackage{graphicx}
\usepackage{algorithm,algorithmic}
\usepackage{hyperref}

\usepackage{comment}
\usepackage[font=small,labelfont=bf,textfont=bf]{caption}

\newtheorem{theorem}{Theorem}
\newtheorem{assumption}{Assumption}
\newtheorem{lemma}{Lemma}
\newtheorem{remark}{Remark}

\hypersetup{hidelinks=true}
\usepackage{textcomp}
\def\BibTeX{{\rm B\kern-.05em{\sc i\kern-.025em b}\kern-.08em
    T\kern-.1667em\lower.7ex\hbox{E}\kern-.125emX}}
\begin{document}
\title{Adaptive Control Barrier Functions for Nonlinear Singularly Perturbed Systems}
\author{Prashanth Krishnamurthy and Farshad Khorrami
\thanks{The authors are with Control/Robotics Research Laboratory (CRRL), Dept. of ECE,
  NYU Tandon School of Engg., Brooklyn, NY 11201, USA. {\tt\small
    \{prashanth.krishnamurthy, khorrami\}@nyu.edu}.
This work is supported in part by the New York University Abu Dhabi (NYUAD)
Center for Artificial Intelligence and Robotics, funded by Tamkeen under the
NYUAD Research Institute Award CG010. An earlier
version~\cite{KK26_cbf_singularly_perturbed} of this paper was presented at the
2026 European Control Conference, Reykjavik, Iceland, July 2026. This manuscript is a preprint of the journal version under review (submitted Aug. 2026).}
}

\maketitle

\bstctlcite{IEEEexample:BSTcontrol}
\begin{abstract}
Control barrier functions (CBFs) have emerged as a flexible and efficient approach for enforcing dynamic safety constraints. In this paper, we consider a class of uncertain nonlinear systems with singular perturbations and develop an adaptive CBF framework for such systems to enforce safety constraints on both the slow and fast subsystems in the overall singularly perturbed nonlinear system with uncertain parameters. We first develop a robustness property showing that if an adaptive controller for the slow subsystem satisfies the slow CBF condition with an appropriate safety margin, then for sufficiently small values of the singular perturbation parameter $\varepsilon$, the slow subsystem remains within an $O(\varepsilon)$-neighborhood of its safe set. We then consider the more general case of safety constraints on both the fast and slow subsystems and provide conditions under which joint slow-fast safety is achieved despite the parametric uncertainties. The reduced-order slow model and the slow and fast CBF conditions are evaluated using an online parameter estimate and the parameter mismatch between the estimated and true reduced-order models is addressed via a state-dependent robustness margin in the CBF inequalities. To illustrate the proposed framework, we present simulation studies for two systems with parametric uncertainties: a synthetic two-time-scale fourth-order example system and a unicycle with a fast turn-rate actuator and an obstacle avoidance safety constraint.
\end{abstract}


\section{Introduction}
\IEEEPARstart{C}{BFs}~\cite{prajna2004safety,wieland2007constructive,wieland2007ensuring,xu2016robustness,ames2017control,ames2019control} have attracted significant research interest over the last several years as an efficient and light-weight approach to enforce safety constraints on dynamical systems.
CBFs are typically applied as a safety filter (typically as a Quadratic Program, i.e., QP) operating in conjunction with a nominal controller to efficiently and minimally adapt control commands during system operation to ensure that safety requirements are satisfied.
CBFs are typically formulated via a safe set over which the CBF is non-negative and a set of inequalities that ensures that the safe set remains forward invariant under the closed-loop system dynamics.
CBFs are flexible and can accommodate a wide range of state and input constraints and scale across a wide variety of systems such as robotic systems, electromechanical systems, and automotive systems and can be applied to ensure safety properties such as obstacle avoidance, state safe limit enforcement, and various application-specific constraints~\cite{glotfelter2019hybrid,garg2024advances,wei2024diffocclusion,wei2025collision}.

A core challenge in application of CBFs is their construction, especially for higher-dimensional nonlinear systems and if non-conservative CBFs are to be designed. CBFs and their construction and analysis have been addressed extensively in the literature over the last two decades~\cite{prajna2004safety,wieland2007constructive,wieland2007ensuring,xu2016robustness,ames2017control,ames2019control,emam2022data,molnar2023composing,goncalves2024safe}. CBF constructions for combinations/compositions of systems have also been considered~\cite{nejati2022compositional,yang2025control}.
While initial CBF literature addressed systems with relative degree one, exponential CBF (ECBF) or high-order CBF methods~\cite{nguyen2016exponential,xiao2022highorder,wei2025collision} have been developed to handle systems with relative degree greater than one.
Adaptive CBF designs have also been developed to address uncertain systems to attain safety guarantees inspite of parametric uncertainties~\cite{lopez2021robust,wang2024adaptive,shen2025adaptive}. Adaptive CBF methods have also been addressed to handle time variations in control input bounds~\cite{xiao2022adaptive}.
Moving beyond hand-crafted CBFs, data-driven approaches also reduce the need for complete knowledge of the system dynamics and enable on-line learning of the control barrier function~\cite{jagtap2020control,khan2022gaussian,dai2022learning,zhang2023learning,xiao2023barriernet}.
Since CBF-based filtering of the control input signal addresses essentially a local optimization problems, methods such as Hessian-based CBF modifications~\cite{reis2021control} and addition of circulation-based inequalities~\cite{goncalves2024control} have been developed to mitigate possible resulting local equilibria in the application of CBFs.
However, despite the wide literature in these directions, a noteworthy gap in the literature is the lack of analysis and design approaches for CBFs for singularly perturbed systems, which represent a widely prevalent structure in real-world systems. An initial step in this direction was taken in our conference paper~\cite{KK26_cbf_singularly_perturbed}, where a CBF-based controller was developed for a class of singularly perturbed systems with completely known dynamics.

Singularly perturbed systems~\cite{kokotovic1999singular,naidu2001singular,khalil2002nonlinear,narang2014nonlinear} appear in a wide variety of real-world systems, including electromechanical systems, robots and autonomous vehicles, chemical processes, and biological systems, and are characterized by the property that they are comprised of subsystems that evolve on widely separated time scales, commonly referred to as `````slow'' and ``fast'' subsystems. For example, in electromechanical systems, the electrical dynamics typically evolve on a much faster time scale than the mechanical dynamics. Similarly, in robotic systems, the robot mechanical dynamics evolve on a much slower time scale than the actuator dynamics. Singular perturbation structures can arise due to presence of subsystems with widely different physical properties, subsystems with different designs of high-gain feedback loops that induce time scale separation, or due to the presence of physical components with widely different time constants.

Classical singular perturbation based controllers exploit the time-scale separation to simplify controller design by decomposing the system into a reduced-order (slow) model and a boundary-layer (fast) subsystem and separately designing controllers for these lower-dimensional subsystems~\cite{kokotovic1999singular,naidu2001singular,khalil2002nonlinear,narang2014nonlinear}, and then showing that a combination of these controllers can achieve desired stability properties for the overall system. However, such singular perturbation based controller designs have not so far been explored in the context of CBFs for enforcing safety constraints.
There are both interesting questions as to whether such a decomposition-based approach can be applied to CBF design for singularly perturbed systems and also practical benefits in terms of enabling reduction of the complexity of CBF design by decomposing the system into lower-dimensional subsystems.

In particular, when applying a nominal CBF-based controller designed using just the reduced-order (slow) system to the full singularly‐perturbed system, it is not obvious whether the desired safety properties are preserved in the presence of the ignored fast dynamics. Also, if safety constraints are desired on both the slow and fast subsystems, it is not clear whether CBF conditions can be designed separately for the slow and fast subsystems and then applied simultaneously to ensure safety of the overall system. Both these questions are addressed in this paper considering a general class of uncertain nonlinear singularly perturbed systems. In this paper, we build on our earlier conference paper~\cite{KK26_cbf_singularly_perturbed} and in comparison, provide a more detailed development of the methodology, extend the results to the case of singularly perturbed systems with parametric uncertainties, and develop an adaptive CBF framework for such systems.

We show that under a suitable boundary‐layer exponential stability condition, an adaptive CBF-based controller designed for the uncertain slow subsystem (with  on-line adaptive parameter estimates used in place of the uncertain parameters in the slow manifold formulation) ensures safety in the overall singularly perturbed system for sufficiently small values of the singular perturbation parameter $\varepsilon$. We also show that explicit safety constraints formulated separately for the slow and fast states can be applied together to obtain joint slow-fast safety in the overall system inspite of the parametric uncertainties.

While the conference version~\cite{KK26_cbf_singularly_perturbed} considered singularly perturbed systems with completely known dynamics, we develop an adaptive CBF framework in this paper in which the reduced-order model and CBF conditions are evaluated using an online parameter estimate $\hat\theta$, with the resulting parameter mismatch addressed by state-dependent robustness margins integrated into the CBF inequalities so that the above safety guarantees are maintained despite the parametric uncertainties.
In particular, a key contribution of this paper is a detailed analysis of the complex interplay between the appearance of the unknown parameters in the singularly perturbed uncertain system dynamics, the online parameter adaptation, and the resulting effects on the slow manifold equations, the reduced-order model dynamics, and the CBF constructions with state-dependent robustness margins.
We also provide two simulation examples for singularly perturbed systems with uncertain parameters to illustrate the proposed framework, namely a synthetic two-time-scale fourth-order example system and a unicycle with a fast turn-rate actuator and including obstacle avoidance as a safety constraint.

\noindent\textbf{Paper Outline:} Section~\ref{sec:notation} summarizes the considered uncertain singularly perturbed system structure and CBF preliminaries. The reduced-model practical safety result for slow subsystem safety constraints is presented in Section~\ref{sec:slow_only}. In Section~\ref{sec:slow_fast}, the more general case of joint slow--fast safety is addressed with a detailed analysis of the safety properties of the overall system. Two illustrative examples are provided in Section~\ref{sec:examples} along with simulation studies. Section~\ref{sec:conclusion} concludes the paper along with a summary of future directions.

\section{System Structure and Preliminaries}
\label{sec:notation}
In this section, an adaptive reduced-order representation of nonlinear singularly perturbed systems is developed. Unlike the formulation in \cite{KK26_cbf_singularly_perturbed}, where the system dynamics are assumed to be completely known, this paper considers the case where the system contains unknown constant parameters that must be estimated online while simultaneously enforcing state safety constraints.

\noindent{\bf System model:}
Consider a class of uncertain singularly perturbed nonlinear systems with slow and fast dynamics given by:
\begin{align}
\dot{x} &= f(x,z,\theta) + g(x,z,\theta)u \label{eq:slow}\\
\varepsilon \dot{z} &= r(x,z,\theta) + s(x,z,\theta)u \label{eq:fast}
\end{align}
where $x\in\mathcal{X}\subset\mathbb{R}^{n_x}$ is the state of the slow subsystem, $z\in\mathcal{Z}\subset\mathbb{R}^{n_z}$ is the state of the fast subsystem, $u\in\mathcal{U}\subset\mathbb{R}^m$ is the control input, $0<\varepsilon\ll1$ is the small parameter, and $\theta\in\Theta\subset\mathbb{R}^{n_\theta}$ is the unknown constant parameter vector appearing in the system dynamics, with $\Theta$ being a known compact convex set and with the bound $\|\theta\|\le\theta_{\max}$ being available.
It is assumed that due to physical limits, the control input is constrained to $\mathcal{U}$, which is a compact set denoting the set of all admissible control inputs, and that the rate of control input change $\dot u$ is constrained such that $\|\dot u\|\le \overline U$.
Also, $\mathcal{X}$ and $\mathcal{Z}$ are compact sets denoting the domains of slow and fast subsystem states, respectively. As in the standard formulation in CBF literature~\cite{ames2017control,ames2019control}, a nominal controller is utilized to ensure basic stability of the system and the CBF-based controller acts as a filter or supervisory controller to ensure safety. Hence, keeping the states $x$ and $z$ within the bounds of sets $\mathcal{X}$ and $\mathcal{Z}$, respectively, is assumed to be ensured by the nominal controller, and is not explicitly considered in this paper for simplicity and brevity (i.e., the set $\mathcal{X}\times\mathcal{Z}$ is assumed to be invariant under the nominal controller). Instead, we focus on the task of the CBF-based controller, which is to enforce the safety constraints discussed below.
Also, in combination with the nominal controller, it is assumed that an estimator is available that generates an estimate $\hat\theta\in\Theta$ of the unknown parameter vector $\theta$ online, with the estimation error $\tilde\theta=\theta-\hat\theta$, which, with the adaptation mechanism, is assumed to satisfy a time-varying bound
\begin{align}
  \|\tilde\theta(t)\|\le\beta_\theta(t)
  \label{eq:tilde_theta_bound}
\end{align}
for $t\ge0$ with $\beta_\theta$ being a known non-negative function. A trivial bound which is guaranteed for this purpose is $\beta_\theta(t)=2\theta_{max}$ simply by the bounded set $\Theta$.
For each fixed $x$, the \emph{slow manifold} (fast equilibrium) is defined as a function of $x$, $u$, and the parameter $\theta$ as the solution $\phi(x,u,\theta)$ to the equation
\begin{equation}
 r(x,\phi(x,u,\theta),\theta) + s(x,\phi(x,u,\theta),\theta)\,u = 0.
\label{eq:slow_manifold}
\end{equation}

\begin{assumption}[Local Lipschitz Condition]\label{asm:smooth}
The functions $f,g,r$, and $s$ are continuously differentiable with respect to $(x,z,\theta)$ and locally Lipschitz uniformly over the compact parameter set $\Theta$. For all piecewise continuous control signals $u(.)$, the solutions of~\eqref{eq:slow}-\eqref{eq:fast} exist and are unique for all $t\ge0$.
\end{assumption}

\begin{assumption}[Bounded Slow Manifold Derivatives]\label{asm:phi_bounded}
The function $\phi(x,u,\theta)$ is continuously differentiable in $(x,u,\theta)$ and constants $L_{\phi,x}, L_{\phi,u}, L_{\phi,\theta} > 0$ exist such that for all $(x,u,\theta) \in \mathcal{X} \times \mathcal{U} \times \Theta$:
\begin{align}
  \left\| \frac{\partial \phi}{\partial x} \right\| \le L_{\phi,x}, \quad
  \left\| \frac{\partial \phi}{\partial u} \right\| \le L_{\phi,u}, \quad
  \left\| \frac{\partial \phi}{\partial \theta} \right\| \le L_{\phi,\theta}.
\end{align}
\end{assumption}

Note that under Assumptions \ref{asm:smooth}-\ref{asm:phi_bounded}, a simple application of the mean value theorem implies that the following inequality is satisfied for all $(x,u,\theta) \in \mathcal{X} \times \mathcal{U} \times \Theta$:
\begin{align}
\|\phi(x,u,\theta)-\phi(x,u,\hat\theta)\| \leq L_\theta\|\tilde\theta\|.
\end{align}

\noindent{\bf Reduced-order model:}
With the slow manifold defined as $z=\phi(x,u,\theta)$, the \emph{reduced-order system} obtained by replacing $z$ in the dynamics of $x$ by the slow manifold approximation of $z$ is given by
\begin{align}
\dot{x} &= \bar f(x,u,\theta) + \bar g(x,u,\theta)u
\label{eq:reduced}
\end{align}
\begin{align}
  \bar f(x,u,\theta) &= f(x,\phi(x,u,\theta),\theta)
  \label{eq:bar_f_defn}
  \\
  \bar g(x,u,\theta) &= g(x,\phi(x,u,\theta),\theta).
                \label{eq:bar_g_defn}
\end{align}
Since the parameter vector $\theta$ is unknown, the slow manifold and the reduced-order model are not directly available for controller design. Instead, in the adaptive CBF framework developed in this paper, these quantities are evaluated using the estimate $\hat\theta$ generated by the adaptation law, with the resulting parameter estimation error accommodated within the CBF conditions in the subsequent sections. In particular, since the true parameter vector is not available, the fast state $z$ in the slow subsystem dynamics \eqref{eq:slow} is replaced by the slow manifold approximation evaluated at the parameter estimate, i.e., $z=\phi(x,u,\hat\theta)$, resulting in the estimated reduced-order model
\begin{align}
\dot{x} &= \hat f(x,u,\hat\theta) + \hat g(x,u,\hat\theta)u
\label{eq:reduced_hat}
\end{align}
\begin{align}
  \hat f(x,u,\hat\theta) &= f(x,\phi(x,u,\hat\theta),\hat\theta)
  \label{eq:hat_f_defn}
  \\
  \hat g(x,u,\hat\theta) &= g(x,\phi(x,u,\hat\theta),\hat\theta).
                \label{eq:hat_g_defn}
\end{align}
In terms of the functions $\hat f$ and $\hat g$ defined in \eqref{eq:hat_f_defn} and \eqref{eq:hat_g_defn}, the dynamics of the slow subsystem can be written as
\begin{align}
\dot x &= \hat f(x,u,\hat\theta) + \hat g(x,u,\hat\theta)u + \Delta_{sp} + \Delta_{ad}
\label{eq:xdot_with_perturbations}
\end{align}
where $\Delta_{sp}$ denotes the singular perturbation error generated by replacing the fast subsystem with its slow-manifold approximation, and $\Delta_{ad}$ denotes the error due to parameter estimation. With
\begin{align}
\tilde f(x,z,\theta,u) &= f(x,z,\theta) - \bar f(x,u,\theta)
\label{eq:tilde_f_defn}
\\
\tilde g(x,z,\theta,u) &= g(x,z,\theta) - \bar g(x,u,\theta),
\label{eq:tilde_g_defn}
\end{align}
$\Delta_{sp}$ is defined as
\begin{align}
\Delta_{sp} = \tilde f(x,z,\theta,u) + \tilde g(x,z,\theta,u)u.
\end{align}
Also,
\begin{align}
\Delta_{ad} &= f(x,\phi(x,u,\theta),\theta)-\hat f(x,u,\hat\theta) \nonumber\\
&\quad + (g(x,\phi(x,u,\theta),\theta) - \hat g(x,u,\hat\theta))u.
\end{align}
Note that $\Delta_{sp}$ is zero on the slow manifold where $z=\phi(x,u,\theta)$, and $\Delta_{ad}$ is zero when the parameter estimate is equal to the true parameter, i.e., when $\hat\theta=\theta$.

\noindent{\bf Slow subsystem safety constraint and CBF:}
The slow-state safety constraint is modeled using a continuously differentiable function $h_x:\mathbb{R}^{n_x}\to\mathbb{R}$ and the requirement that $h_x(x)\ge0$ along all trajectories of the system, i.e., the safe set is defined as $\mathcal{C}_x := \{x\in\mathbb{R}^{n_x}\mid h_x(x)\ge0\}$.
Define
\begin{align}
  \Delta_{h_x} &= L_{\bar f(\cdot;u)}h_x(x) + L_{\bar g(\cdot;u)}h_x(x)u \nonumber\\
  &\quad - L_{\hat f(\cdot;u)}h_x(x) - L_{\hat g(\cdot;u)}h_x(x)u,
\label{eq:Delta_hx}
\end{align}
where $L_{\bar f(\cdot;u)} h_x(x)=\nabla h_x(x)^\top \bar f(x,u,\theta)$, $L_{\bar g(\cdot;u)} h_x(x)=\nabla h_x(x)^\top \bar g(x,u,\theta)$, $L_{\hat f(\cdot;u)} h_x(x)=\nabla h_x(x)^\top \hat f(x,u,\hat\theta)$, and $L_{\hat g(\cdot;u)} h_x(x)=\nabla h_x(x)^\top \hat g(x,u,\hat\theta)$, with $\nabla h_x(x)$ denoting the transpose of $\frac{\partial h_x}{\partial x}$.
In \eqref{eq:Delta_hx}, we write $L_{\bar f(\cdot;u)} h_x(x)$, $L_{\bar g(\cdot;u)} h_x(x)$, etc., to emphasize the $u$-dependence through $\phi(x,u,\hat\theta)$; for brevity in later equations, we will just write $L_{\bar f}h_x(x)$, $L_{\bar g}h_x(x)$, etc., for these quantities.
With $\Delta_{h_x}$ defined in \eqref{eq:Delta_hx},
it can be seen from the local Lipschitz continuity of $f$ and $g$ (Assumption~\ref{asm:smooth}), the boundedness of slow-manifold derivatives (Assumption~\ref{asm:phi_bounded}), the compact control set $\mathcal{U}$, and the parameter estimation error bound in \eqref{eq:tilde_theta_bound} that a bound of the form
\begin{align}
  \|\Delta_{h_x}\| \leq \sigma_x(x)\beta_\theta(t)
  \label{eq:Delta_hx_bound}
\end{align}
holds with $\sigma_x(x) = \|\nabla h_x(x)\|\,\sigma_{x0}$ for a positive constant $\sigma_{x0}$ depending on the Lipschitz constants of $f, g, \phi$ and the bound on control input set $\mathcal{U}$.

The function $h_x$ is said to be an adaptive CBF\footnote{For simplicity, we focus on the case where the relative degree from the CBF to the control input is 1. However, the proposed framework can be applied to cases with higher relative degree using higher-order CBF or exponential CBF (ECBF) formulations developed in the literature\cite{nguyen2016exponential,xiao2022highorder}. Example 3 in Section~\ref{sec:examples} provides a practical example of such an application.} for the estimated reduced-order system \eqref{eq:reduced_hat} with margin $\gamma_x\geq 0$ if an extended class-$\mathcal{K}$ function $\alpha_x$ exists that is locally Lipschitz around the origin such that\cite{ames2017control,ames2019control}
\begin{align}
\sup_{u\in\mathcal{U}}\big[ &L_{\hat f}h_x(x) \!+\! L_{\hat g}h_x(x)u \!+\! \alpha_x(h_x(x)) \big] \ge \gamma_x+\sigma_x(x)\beta_\theta(t).
\label{eq:cbf_cond}
\end{align}
In \eqref{eq:cbf_cond}, while $\gamma_x$ is a safety margin included for ``robustness,'' the term $\sigma_x(x)\beta_\theta(t)$ is a safety margin addressing the parameter adaptation and is time-varying and could reduce over time depending on the parameter estimator properties.
Since the CBF condition \eqref{eq:cbf_cond} is evaluated using the parameter estimate $\hat\theta$ in place of the true parameter $\theta$, the difference between the estimated and true Lie derivatives is bounded in terms of the parameter estimation error $\|\hat\theta-\theta\|$ under Lipschitz conditions on the system dynamics with respect to $\theta$, and is accommodated within the adaptive CBF analysis developed in the subsequent sections.

\noindent{\bf Fast subsystem safety constraint and CBF:}
Analogous to the slow subsystem safety constraint defined above, we consider the fast subsystem safety constraint as modeled by a continuously differentiable function $h_z:\mathbb{R}^{n_x}\times\mathbb{R}^{n_z}\to\mathbb{R}$
and the requirement that $h_z(x,z)\ge0$ along all trajectories of the system, i.e.,
the $x$-dependent safe set for the fast subsystem state is defined as
 $\mathcal{C}_z := \{z\in\mathbb{R}^{n_z}\mid h_z(x,z)\ge0\}$.
Define
\begin{align}
  \Delta_{h_z} &= L_{r,z}h_z(x,z) + L_{s,z}h_z(x,z)u \nonumber\\
  &\quad - L_{\hat r,z}h_z(x,z) - L_{\hat s,z}h_z(x,z)u
\label{eq:Delta_hz}
\end{align}
where $L_{r,z} h_z(x,z)=\nabla_z h_z(x,z)^\top r(x,z,\theta)$, $L_{s,z} h_z(x,z)=\nabla_z h_z(x,z)^\top s(x,z,\theta)$, $L_{\hat r,z} h_z(x,z)=\nabla_z h_z(x,z)^\top \hat r(x,z,\hat\theta)$, and $L_{\hat s,z} h_z(x,z)=\nabla_z h_z(x,z)^\top \hat s(x,z,\hat\theta)$, with $\hat r(x,z,\hat\theta)=r(x,z,\hat\theta)$, $\hat s(x,z,\hat\theta)=s(x,z,\hat\theta)$, and with $\nabla_z h_z(x,z)$ denoting $(\frac{\partial h_z}{\partial z})^\top$.
With $\Delta_{h_z}$ defined in \eqref{eq:Delta_hz},
it can be seen from the local Lipschitz continuity of $r$ and $s$ (Assumption~\ref{asm:smooth}), the compact control set $\mathcal{U}$, and the parameter estimation error bound in \eqref{eq:tilde_theta_bound} that a bound of the form
\begin{align}
  \|\Delta_{h_z}\| \leq \sigma_z(x,z)\beta_\theta(t)
  \label{eq:Delta_hz_bound}
\end{align}
holds with $\sigma_z(x,z) = \|\nabla_z h_z(x,z)\|\,\sigma_{z0}$ with $\sigma_{z0}$ being a positive constant that depends on the Lipschitz constants of $r, s$ and the bound on control input set $\mathcal{U}$.

The function $h_z$ is said to be an adaptive CBF for the fast subsystem with margin $\gamma_z\geq 0$ if, with the fast subsystem dynamics \eqref{eq:fast} evaluated at the parameter estimate $\hat\theta$, there exists an extended class-$\mathcal{K}$ function $\alpha_z$ that is locally Lipschitz around the origin such that
 \begin{align}
 &\sup_{u\in\mathcal{U}}\big[ L_{\hat r,z} h_z(x,z) + L_{\hat s,z} h_z(x,z)\,u
 \nonumber\\
 &\qquad\qquad\qquad +\alpha_z(h_z(x,z)) \big] \ge \gamma_z + \sigma_z(x,z)\beta_\theta(t).
 \label{eq:cbf_cond_fast}
 \end{align}
Analogously to the slow subsystem CBF condition \eqref{eq:cbf_cond}, the term $\sigma_z(x,z)\beta_\theta(t)$ in \eqref{eq:cbf_cond_fast} is a time-varying safety margin accommodating parameter estimation errors, while $\gamma_z$ is a margin for robustness.

\section{Robust Slow-State Safety}
\label{sec:slow_only}

In this section, we will first consider the case where only the slow subsystem safety constraint is to be satisfied and show that under the assumptions below, the slow subsystem's safety constraint is satisfied even in the presence of the fast subsystem.
As can be seen below, these technical assumptions are reasonably mild and along the standard lines of the literature on singularly perturbed systems~\cite{khalil2002nonlinear} and easy to check for given systems (see examples in Section~\ref{sec:examples}).
In Section~\ref{sec:slow_fast}, we will consider the general case where both slow and fast subsystem safety constraints are to be satisfied  and show that the slow and fast CBF conditions can be applied simultaneously under appropriate assumptions to obtain joint slow--fast safety.

\begin{assumption}[Boundary-layer stability of fast subsystem\footnote{For simplicity, we consider the case of stable fast subsystem dynamics, which is a common assumption in the singular perturbation literature~\cite{khalil2002nonlinear} and is satisfied in typical real-world systems (see Section~\ref{sec:examples}). Also, in systems where the fast subsystem is directly affected by control inputs and the control law can depend on the fast subsystem's state, the control inputs can also be used to stabilize the fast subsystem, but this case is not considered here for simplicity and brevity.}]\label{asm:faststab}
For each fixed $x$, consider the fast subsystem below obtained with the time scale transformation $\tau = t/\varepsilon$:
\begin{align}
\frac{d\zeta}{d\tau} = r(x,\zeta,\theta) + s(x,\zeta,\theta)u + \delta
\end{align}
with $\delta$ being an exogenous disturbance input.
With $\delta\equiv 0$, this fast subsystem has a unique exponentially stable equilibrium $\zeta=\phi(x,u,\theta)$ where $\phi(x,u,\theta)$ was defined by the slow manifold equation in \eqref{eq:slow_manifold}.
There exist constants $c>0$, $\lambda>0$ such that
\begin{align}
\|\zeta(\tau)-\phi(x,u,\theta)\| &\le c\,e^{-\lambda \tau}\|\zeta(0)-\phi(x,u,\theta)\|
\nonumber\\
&\quad + c\mu(\sup_{\tau_1\in[0,\tau]} \|\delta(\tau_1)\|)
\end{align}
where $\mu$ is a class-$\mathcal{K}$ function that is locally Lipschitz around the origin. For example, for a linear-like system, we would have $\mu(\sup_{\tau_1\in[0,\tau]} \|\delta(\tau_1)\|)=\int_{0}^{\tau} e^{-\lambda (\tau-\tau_2)} \sup_{\tau_1\in[0,\tau]}\|\delta(\tau_1)\| d\tau_2$.
\end{assumption}

\begin{assumption}[CBF feasibility for the slow subsystem safety constraint \eqref{eq:cbf_cond} with estimated reduced-order system \eqref{eq:reduced_hat}]\label{asm:margin}
A $\gamma_x>0$ and an extended class-$\mathcal{K}$ function $\alpha_x$ that is locally Lipschitz around the origin exist such that for each $x \in \mathcal{C}_x$ and parameter estimate $\hat\theta \in \Theta$,
a control value $u\in\mathcal{U}$ can be found to satisfy the inequality
\begin{equation}
L_{\hat f}h_x(x) + L_{\hat g}h_x(x)u \ge -\alpha_x(h_x(x)) + \gamma_x + \sigma_x(x)\beta_\theta(t).
\label{eq:margin}
\end{equation}
\end{assumption}

\begin{assumption}[Slow manifold deviation mismatch bound]\label{asm:mismatch}
  A class-$\mathcal{K}$ function $\rho$ exists that is locally Lipschitz around the origin such that for all $x\in\mathcal{C}_x$, $z\in\mathcal{C}_z$, $u\in\mathcal{U}$, and $\theta\in\Theta$, the following inequality is satisfied:
\begin{align}
\!\!\!\!\!\!\big|\nabla h_x(x)^\top\!(\tilde f(x,z,\theta,u) \!+\! \tilde g(x,z,\theta,u)\,u)\big|  \le \rho(\|z-\phi(x,u,\theta)\|)\!\!\!
\label{eq:mismatch}
\end{align}
where $\tilde f(x,z,\theta,u)$ and $\tilde g(x,z,\theta,u)$ are defined in \eqref{eq:tilde_f_defn} and \eqref{eq:tilde_g_defn}, respectively.
Note that from the definitions of $\tilde f(x,z,\theta,u)$ and $\tilde g(x,z,\theta,u)$, it is seen that on the slow manifold where $z=\phi(x,u,\theta)$, we have $\tilde f(x,z,\theta,u)=0$ and $\tilde g(x,z,\theta,u)=0$.
\end{assumption}

We first consider the dynamics of the ``tracking error'' of the fast subsystem and provide a bound on the error as a lemma.
The fast subsystem tracking error is defined as $e  =z-\phi(x,u,\theta)$. Differentiating and using~\eqref{eq:slow_manifold}, we obtain
\begin{align}
 \varepsilon \dot e &= r\big(x,\phi(x,u,\theta)+e,\theta\big) + s\big(x,\phi(x,u,\theta)+e,\theta\big)u \nonumber\\
 &\quad - \varepsilon\,\frac{\partial\phi}{\partial x}\dot x - \varepsilon\,\frac{\partial\phi}{\partial u}\dot u.
\end{align}

\begin{lemma}[Bound on boundary-layer tracking error]\label{lem:bl}
Under Assumptions~\ref{asm:smooth}--\ref{asm:faststab}, for any compact time interval $[0,T]$ on which the solution trajectories remain in a compact set, then denoting $\tau=\frac{t}{\varepsilon}$ and $M_x = \sup_{t\in[0,T]}\|\dot x(t)\|$, the inequality below is satisfied for all $t\in[0,T]$:
\begin{equation}
  \|e(t)\| \le c\,e^{-\lambda \tau}\,\|e(0)\| + c\mu(\varepsilon\bar E)
  \label{eq:lemma1}
\end{equation}
where $\bar E=L_{\phi,x} M_x+L_{\phi,u}\overline U$.
In particular, \eqref{eq:lemma1} implies that the error $e$ satisfies the inequality $\|z(t)-\phi(x(t),u(t),\theta)\|\le 2c\mu(\varepsilon \bar E)$ after a short boundary-layer transient (in particular, after $t\ge (\varepsilon/\lambda)\log(\|e(0)\|/\mu(\varepsilon \bar E))$). Note that this transient time is of order $O(\varepsilon\log(1/\varepsilon))$ since $\mu$ is a class-$\mathcal{K}$ function that is locally Lipschitz around the origin.
\end{lemma}
\begin{proof}
By exponential stability of the fast dynamics for any frozen $x$ and parameter vector $\theta\in\Theta$ and input-to-state stable (ISS) property of the boundary-layer system from Assumption~\ref{asm:faststab} with $\delta=- \varepsilon(\partial\phi/\partial x)\dot x - \varepsilon(\partial\phi/\partial u)\dot u$, we obtain
the bound in the lemma since
$\|\partial\phi/\partial x\|\le L_{\phi,x}$, $\|\partial\phi/\partial u\|\le L_{\phi,u}$, $\|\dot u\|\le\bar U$, and $\|\dot x\|\le M_x$ over $[0,T]$. Note that existence of the bound $M_x$ follows from the compactness of the set on which the solution trajectories evolve over $[0,T]$ and the local Lipschitz property of the system dynamics from Assumption~\ref{asm:smooth}.
\end{proof}

We now present the main result on robust slow-state safety in the presence of the fast dynamics in the overall singularly perturbed system.

\begin{theorem}\label{thm:main}
Under Assumptions~\ref{asm:smooth}--\ref{asm:mismatch}, a constant $\varepsilon^\star>0$ exists such that for all $0<\varepsilon\le\varepsilon^\star$, any solution $(x(t),z(t))$ of~\eqref{eq:slow}--\eqref{eq:fast} in closed loop with an adaptive CBF controller satisfying the condition in Assumption~\ref{asm:margin} with margin $\gamma_x$ starting with $x(0)\in\mathcal{C}_x$ satisfies the ``practical safety'' property
\begin{align}
 h_x(x(t)) \ge -\delta_x(\varepsilon) \quad \forall t\ge \underline t(\varepsilon)
\end{align}
where $\delta_x(\varepsilon)=O(\varepsilon)$ and $\underline t(\varepsilon)$ is a function that is $O(\varepsilon\log(1/\varepsilon))$. A sufficient (possibly conservative)
bound for $\varepsilon^\star$ is
\begin{align}
\varepsilon^\star = \frac{1}{\overline E}\mu^{-1}\Big(\frac{\rho^{-1}(\gamma_x)}{2c}\Big).
\label{eq:eps_bound}
\end{align}
Additionally, if $\varepsilon\leq \varepsilon^\star$ with the $\varepsilon^\star$ given in \eqref{eq:eps_bound}, the strict safety property given by the stronger inequality $h_x(x(t))\ge0$ holds for all $t\geq\underline t(\varepsilon)$.
\end{theorem}

\begin{proof}
Under the control law for $u$ from Assumption~\ref{asm:margin}, the actual slow subsystem dynamics from \eqref{eq:slow} can be written as the estimated reduced-order dynamics along with the singular perturbation and parameter-estimation errors as shown in \eqref{eq:xdot_with_perturbations}. Differentiating the slow subsystem's CBF $h_x(x)$ along the trajectories of the system, and using the definition of $\Delta_{h_x}$ in~\eqref{eq:Delta_hx}, we have
\begin{align}
\!\!\!\!\!\dot{h}_x(x)
= L_{\hat f}h_x(x) + L_{\hat g}h_x(x)u + \Delta_{h_x}
\!+\! \nabla h_x(x)^\top\Delta_{sp}(x,z,u).\!\!\!
\end{align}
Using~\eqref{eq:Delta_hx_bound}, \eqref{eq:margin}, and~\eqref{eq:mismatch}, for all $t\in[0,T]$,
\begin{align}
\dot{h}_x(x) \ge -\alpha_x(h_x(x)) + \gamma_x - \rho(\|z-\phi(x,u,\theta)\|).
\end{align}
From Lemma~\ref{lem:bl}, after a short boundary-layer transient time $\underline t(\varepsilon)$ that is $O(\varepsilon\log(1/\varepsilon))$, the quantity $\|z-\phi(x,u,\theta)\|$ is bounded by $2c\mu(\varepsilon \overline E)$, implying that
\begin{align}
\dot{h}_x(x) \ge -\alpha_x(h_x(x)) + \gamma_x - \eta_x(\varepsilon)
\label{eq:dot_hx_bound}
\end{align}
where $\eta_x(\varepsilon)=\rho(2c\mu(\varepsilon \overline E))$. Whenever $\alpha_x(h_x(x)) \leq \gamma_x-\eta_x(\varepsilon)$, then we have $\dot h_x \ge 0$. Since $\alpha_x$ is an extended class-$\mathcal{K}$ function, $\rho$ and $\mu$ are class-$\mathcal{K}$ functions, and $\alpha_x$, $\rho$, and $\mu$ are locally Lipschitz around the origin, this implies that $h_x(x(t))\ge -\delta_x(\varepsilon)$ for $t\geq\underline t(\varepsilon)$, with $\delta_x(\varepsilon)=O(\varepsilon)$.
In particular, considering for example, a linear extended class-$\mathcal K$ function $\alpha_x(h_x)=\kappa_h h_x$, an explicit form can be obtained for $\delta_x(\varepsilon)$ as $\delta_x(\varepsilon)=\eta_x(\varepsilon)/\kappa_h$.
Finally, noting that if $\eta_x(\varepsilon)\leq \gamma_x$, the analysis above reduces to $h_x(x(t))\ge 0$ after the boundary-layer transient, a sufficient (conservative) smallness condition for $\varepsilon^*$ is $\eta_x(\varepsilon^*)\le\gamma_x$, which directly yields~\eqref{eq:eps_bound}.
\end{proof}

\begin{remark}
  \label{rem:strict_safety_slow}
  While Theorem~\ref{thm:main} establishes practical safety of the slow subsystem after a short boundary-layer transient, strict safety of the slow subsystem for all $t\geq0$ is also guaranteed under additional conditions on the initial deviation from the slow manifold and/or the initial slow-state safety margin.
The strict-safety conclusion can be extended from $t\geq\underline t(\varepsilon)$ to all $t\geq0$ in either of the following cases. First, if $\gamma_x>\rho(c\mu(\varepsilon\overline E))$ and
\begin{align}
 \|e(0)\|\leq\bar e_0,
 \qquad
 \bar e_0=\frac{1}{c}\rho^{-1}(\gamma_x)-\mu(\varepsilon\overline E)>0,
 \label{eq:init_error_bound}
\end{align}
then Lemma~\ref{lem:bl} gives $\|e(t)\|\leq\rho^{-1}(\gamma_x)$ for all $t\geq0$. Consequently, the CBF derivative is nonnegative whenever $h_x(x)=0$, including during the boundary-layer transient, implying that strict safety holds for all $t\geq0$.

Second, suppose that $h_x(x(0))\geq h_{\min}>0$, that $\|e(0)\|\leq E_0$ for a known constant $E_0$ (the existence of which follows from the bounds on the states), and that $H_x$ is a uniform bound on $|\dot h_x|$ over the compact set of admissible slow and fast states, inputs, and parameters. Define
\begin{align}
 \varepsilon_{\rm marg}^\star
 = \min\!\left\{
 \frac{1}{\overline E}\mu^{-1}\!\left(\frac{\rho^{-1}(\gamma_x)}{2c}\right),
 \frac{\lambda h_{\min}}{H_x\log^+\!\left(\frac{2cE_0}{\rho^{-1}(\gamma_x)}\right)}
 \right\},
 \label{eq:eps_margin_bound}
\end{align}
where $\log^+(y)=\max\{\log(y),0\}$ and the second term is interpreted as $+\infty$ when its denominator is zero. Then, for every $0<\varepsilon\leq\varepsilon_{\rm marg}^\star$, strict safety holds for all $t\geq0$, independently of the particular initial error satisfying $\|e(0)\|\leq E_0$. Indeed, with
\begin{align}
 t_{\rm bl}=\frac{\varepsilon}{\lambda}\log^+\!\left(\frac{2cE_0}{\rho^{-1}(\gamma_x)}\right),
\end{align}
Lemma~\ref{lem:bl} gives $\|e(t)\|\leq \rho^{-1}(\gamma_x)$ for $t\geq t_{\rm bl}$, while the possible decrease in $h_x$ before $t_{\rm bl}$ is at most $H_xt_{\rm bl}\leq h_{\min}$. The CBF derivative is then nonnegative on the boundary after $t_{\rm bl}$, so the safe set remains forward invariant.
\end{remark}

\begin{remark}
For a linear extended class-$\mathcal K$ function $\alpha_x(h_x)=\kappa_h h_x$, it was seen in Theorem~\ref{thm:main} and its proof that the maximum deviation (``tube size'') of the practical safety of the slow subsystem after the initial transient can be written explicitly as
$\delta_x(\varepsilon) = \rho(2c\mu(\varepsilon \overline E))/\kappa_h$.
The parameter-estimation error does not appear as an additional term in this tube size because it is compensated by the adaptive margin $\sigma_x(x)\beta_\theta(t)$ in the CBF condition. The sufficient smallness condition ensuring nonnegativity of $h_x(x)$ after the initial transient is given by~\eqref{eq:eps_bound}.
Noting the definition of $\overline E$ in Lemma~\ref{lem:bl}, the estimate for $\varepsilon^*$ illustrates the trade-offs related to the choice of the CBF margin and the bounds on the slow-manifold derivatives.
The estimate $\varepsilon^*$ becomes larger (less conservative) with a larger CBF margin and smaller (more conservative) with larger slow-manifold derivative bounds $L_{\phi,x}$ and $L_{\phi,u}$.
\end{remark}

\begin{remark}
The dependence on $L_{\phi,u}$ via $\overline E$ appearing in the constructions of $\delta_x(\varepsilon)$ and $\varepsilon^*$ in the proof of Theorem~\ref{thm:main} captures how the dependence of the slow manifold on the control input influences the safety properties through $\phi(x,u,\theta)$. In the special case that $s(x,z,\theta)\equiv0$, the fast dynamics do not directly depend on the control input, so $\partial\phi/\partial u\equiv0$, $L_{\phi,u}=0$, and $\overline E=L_{\phi,x}M_x$. If both $L_{\phi,x}$ and $L_{\phi,u}$ are zero (weak coupling between the slow and fast subsystems), then $\overline E=0$, the fast-state tracking error has no $O(\varepsilon)$ forcing from $\dot x$ or $\dot u$, and the reduced-order adaptive CBF condition can enforce slow-state safety without a small-$\varepsilon$ restriction arising from this tracking-error bound, i.e., arbitrarily large $\varepsilon$ can be handled and the fast dynamics can be completely neglected. In the general case with non-negligible coupling between the slow and fast subsystems, including direct control dependence in the fast dynamics, Theorem~\ref{thm:main} provides a general characterization of the resulting slow-state safety properties in the presence of the coupled fast dynamics.
\end{remark}

\section{Joint Slow-Fast Safety}
\label{sec:slow_fast}

While Section~\ref{sec:slow_only} showed that slow subsystem safety properties are maintained even in the presence of the fast dynamics, we now consider the more general case where we have safety constraints on both slow states $x$ and fast states $z$. We will show below that in this general case, joint slow-fast safety can be attained by using slow and fast CBF conditions simultaneously under appropriate conditions for the interaction between slow and fast dynamics. We consider the joint safe set $\mathcal{C}_{xz}=\mathcal{C}_x\times\mathcal{C}_z$ and introduce the following assumptions along with the Assumptions \ref{asm:smooth}-\ref{asm:mismatch} from Section~\ref{sec:slow_only}.

\begin{assumption}[Joint adaptive CBF feasibility and slow-fast compatibility]\label{asm:fastcbf}
  Positive constants $\gamma_x$ and $\gamma_z$ and extended class-$\mathcal{K}$ functions
  $\alpha_x$ and $\alpha_z$ that are locally Lipschitz around the origin exist such that for each
  $(x,z)\in\mathcal{C}_{xz}$, parameter estimate $\hat\theta\in\Theta$, and time $t\geq0$,
  a control value $u\in\mathcal{U}$ can be found to satisfy the following properties:
  \begin{itemize}
    \item the adaptive slow CBF inequality~\eqref{eq:margin}, which is based on the reduced-order slow subsystem dynamics;
    \item the adaptive frozen-$x$ fast CBF inequality
\begin{align}
&L_{\hat r,z} h_z(x,z) + L_{\hat s,z} h_z(x,z)\,u
\ge -\alpha_z(h_z(x,z)) \nonumber\\
&\qquad\qquad\qquad\quad\qquad\qquad + \gamma_z
  + \sigma_z(x,z)\beta_\theta(t)
\label{eq:cbf_fast}
\end{align}
    where $\hat r$ and $\hat s$ are evaluated at $\hat\theta$ as in~\eqref{eq:Delta_hz}
    \item the robust slow-fast compatibility condition
    \begin{align}
      h_z\big(x,\phi(x,u,\vartheta)\big)\geq0
      \quad\text{for all }\vartheta\in\Theta_\beta(t)
      \label{eq:fast_slow_compatibility}
    \end{align}
    where $\Theta_\beta(t)=\{\vartheta\in\Theta\mid \|\vartheta-\hat\theta\|\leq\beta_\theta(t)\}$.
    Since the fast state $z$ remains close to the true slow manifold after an initial transient (by Lemma~\ref{lem:bl}), this requirement ensures that the slow and fast safe set constraints are not contradictory for any parameter value consistent with the estimation bound.
\end{itemize}
\end{assumption}

\begin{assumption}[Bound on fast CBF gradient w.r.t. slow state\footnote{This assumption which characterizes the dependence of the fast CBF on the slow state is reasonably mild. Indeed, in many real-world cases, the safety constraints for the fast subsystem do not explicitly depend on the slow state variables, in which case $\frac{\partial h_z}{\partial x}=0$ trivially satisfying the assumption.}]\label{asm:cbf_gradient_bound}
A positive constant $K_z$ exists such that for all $(x,z)$, the inequality
$|\partial h_z/\partial x| \le K_z$ is satisfied.
\end{assumption}

\begin{theorem}[Joint Slow-Fast Safety]\label{thm:joint}
Under Assumptions~\ref{asm:smooth}-\ref{asm:cbf_gradient_bound}, positive constants $\varepsilon_x^\star$ and $\varepsilon_z^\star>0$ exist such that for all
 $0<\varepsilon \le \min\{\varepsilon_x^\star,\varepsilon_z^\star\}$,
any solution $(x(t),z(t))$ of~\eqref{eq:slow}-\eqref{eq:fast} in closed loop with a controller satisfying the conditions in Assumptions~\ref{asm:margin} and \ref{asm:fastcbf} starting with $x(0)\in\mathcal{C}_x$ and $z(0)\in\mathcal{C}_z$ satisfies the ``practical joint slow-fast safety'' property
\begin{align}
 h_x(x(t)) \ge -\delta_x(\varepsilon) \quad\forall t\ge \underline t(\varepsilon) \\
 h_z(x(t),z(t)) \ge -\delta_z(\varepsilon) \quad\forall t\ge 0
\end{align}
where $\delta_x(\varepsilon)$ and $\delta_z(\varepsilon)$ are functions that are $O(\varepsilon)$ and $\underline t(\varepsilon)$ is a function that is $O(\varepsilon\log(1/\varepsilon))$. Furthermore, sufficient (possibly conservative) bounds for $\varepsilon_x^\star$ and $\varepsilon_z^\star$ are
\begin{align}
  \varepsilon_x^\star &=
\frac{1}{\overline E}\mu^{-1}\Big(\frac{\rho^{-1}(\gamma_x)}{2c}\Big)
                         \ \ , \ \
\varepsilon_z^\star = \frac{\gamma_z}{K_z M_x}. \label{eq:epsz}
\end{align}
Additionally, if $\varepsilon\leq \min\{\varepsilon_x^\star,\varepsilon_z^\star\}$ with $\varepsilon_x^\star$ and $\varepsilon_z^\star$ given in \eqref{eq:epsz}, the strict joint slow-fast safety properties given by the stronger inequalities $h_x(x(t))\ge0$ and $h_z(z(t))\ge0$ hold for all $t\ge \underline t(\varepsilon)$ and $t\ge0$, respectively.
\end{theorem}

\begin{proof}
Since Assumptions~\ref{asm:smooth}--\ref{asm:mismatch} are inherited from Theorem~\ref{thm:main}, the slow-state CBF properties follow from that theorem.
To analyze properties of the fast CBF, differentiate $h_z$ along~\eqref{eq:fast}:
\begin{align}
\varepsilon \dot{h}_z
&= L_{\hat r,z}h_z(x,z) + L_{\hat s,z}h_z(x,z)u
 + \Delta_{h_z} \nonumber\\
&\quad + \varepsilon\frac{\partial h_z}{\partial x}\dot x \\
&\ge -\alpha_z(h_z(x,z)) + \gamma_z
 + \varepsilon\frac{\partial h_z}{\partial x}\dot x
\end{align}
where the adaptive fast CBF condition in Assumption~\ref{asm:fastcbf} and the bound on $\Delta_{h_z}$ from~\eqref{eq:Delta_hz_bound} were used in writing the last inequality. Using Assumption~\ref{asm:cbf_gradient_bound} and the bound $\|\dot x\|\leq M_x$ over $[0,T]$ as in Lemma~\ref{lem:bl}, and dividing throughout by $\varepsilon$, we have
\begin{align}
\dot{h}_z
&\geq -\frac{1}{\varepsilon}\alpha_z(h_z(x,z)) + \frac{1}{\varepsilon}\gamma_z - K_z M_x.
\end{align}
Defining $\eta_z(\varepsilon)=K_z M_x\varepsilon$, we see that whenever $\alpha_z(h_z(x,z))\leq\gamma_z-\eta_z(\varepsilon)$ then $\dot h_z \ge0$.
Since $\alpha_z$ is an extended class-$\mathcal K$ function that is locally Lipschitz around the origin, this implies that $h_z(x(t),z(t))\ge -\delta_z(\varepsilon)$ for all $t\in[0,T]$, with $\delta_z(\varepsilon)=O(\varepsilon)$. In particular, considering for example, a linear extended class-$\mathcal K$ function $\alpha_z(h_z)=\kappa_z h_z$, we can write explicitly $\delta_z(\varepsilon)=\eta_z(\varepsilon)/\kappa_z$.
Hence, the fast state $z$ remains within an $O(\varepsilon)$ neighborhood of $\mathcal{C}_z$.
Furthermore, using Lemma 1, we know that after the short boundary-layer transient time $\underline t(\varepsilon)$ which is $O(\varepsilon\log(1/\varepsilon))$, the inequality $\|z-\phi(x,u)\| \leq 2c\mu(\varepsilon \bar E)$ is satisfied, i.e., the fast state remains within $O(\varepsilon)$ of the slow manifold.
Also, noting that if $\eta_z(\varepsilon) \leq \gamma_z$, the analysis above reduces to $h_z(x(t),z(t))\geq 0$ for all time, a sufficient (conservative) smallness condition for $\varepsilon_z^*$ is $\eta_z(\varepsilon^*)\leq \gamma_z$, yielding \eqref{eq:epsz}.
 Combining the bound $\varepsilon^*$ from Theorem~\ref{thm:main} and the above bound $\varepsilon_z^*$, we see that the statement of the theorem follows.
\end{proof}

\begin{remark}
  Analogous to Remark~\ref{rem:strict_safety_slow}, the strict joint slow-fast safety conclusion can be extended from $t\geq\underline t(\varepsilon)$ to all $t\geq0$ by combining the corresponding initial-time conditions for the slow and fast subsystems. In particular, it is sufficient that the slow subsystem satisfy either of the conditions in Remark~\ref{rem:strict_safety_slow} and that $\varepsilon\leq\varepsilon_z^\star$ with $\varepsilon_z^\star$ given in \eqref{eq:epsz}.
\end{remark}

\begin{remark}
  \label{rem:cbf_projection}
  As discussed in Section~\ref{sec:notation}, the typical application of a CBF-based controller is as a safety filter in conjunction with a nominal controller (or ``performance controller'') that handles stabilization and application-specific objectives such as trajectory tracking.
  In implementation, the adaptive CBF constraints are enforced via a QP \cite{ames2017control,ames2019control} that is formulated to find the control input closest to the instantaneous nominal controller value $u_{\mathrm{nom}}$ while satisfying the estimated CBF constraints and their parameter-uncertainty margins. To achieve the joint slow-fast safety properties in Theorem~\ref{thm:joint}, the safety filter can be implemented by solving the CBF-based QP below at each time step:
\begin{align}
 u &= \arg\min_{u\in\mathcal{U}} \|u-u_{\mathrm{nom}}\|^2 \\
 \text{s.t.}\quad & L_{\hat f}h_x(x)+ L_{\hat g}h_x(x)\,u \\
 &\qquad\qquad\ge -\alpha_x(h_x(x)) +\gamma_x+\sigma_x(x)\beta_\theta(t)\\
 & L_{\hat r,z} h_z(x,z) + L_{\hat s,z} h_z(x,z)\,u \\
 &\qquad\qquad\ge -\alpha_z(h_z(x,z)) +\gamma_z+\sigma_z(x,z)\beta_\theta(t).
\label{eq:u_opt}
\end{align}
This adaptive CBF QP enforces safety for both the slow and fast subsystems simultaneously, using the parameter estimate and the corresponding uncertainty margins. Additionally, at each time step, the robust compatibility condition $h_z(x,\phi(x,u,\vartheta))\geq0$ is verified for all $\vartheta\in\Theta_\beta(t)$, as required by~\eqref{eq:fast_slow_compatibility} in Assumption~\ref{asm:fastcbf}. In practice, this is a mild condition as long as the slow and fast safe set constraints are not contradictory, and can be simply verified separately at each time step instead of requiring it formally in the CBF-QP, but can be incorporated into the optimization problem  \eqref{eq:u_opt} as a constraint on $u$ if needed.
\end{remark}

\begin{remark}
Due to the $u$-dependence in the estimated slow manifold $\phi(x,u,\hat\theta)$, the estimated reduced Lie derivatives $L_{\hat f}h_x$ and $L_{\hat g}h_x$ inherit an implicit $u$-dependence. Hence, the adaptive CBF inequality for the estimated reduced-order system can, in general, be non-affine in $u$. In practice, a few approaches can be used to deal with this: (i) obtaining an affine approximation by replacing the input dependence in $\hat f$ and $\hat g$ with the previous control value (i.e., from the previous time step), (ii) computing a linearization for the input dependence of $L_{\hat f}h_x+L_{\hat g}h_xu$, or (iii) computing a conservative lower bound on $L_{\hat f}h_x+L_{\hat g}h_xu$ in an input-affine form and correspondingly tightening the margin $\gamma_x$ to accommodate a bound on the neglected higher-order terms. Furthermore, note that in cases where $g$ in the slow system dynamics \eqref{eq:slow} does not depend on $z$, the term $\hat g$ and $L_{\hat g}h_x$ will not introduce any additional input dependence through the slow manifold. Also, $\phi(x,u,\hat\theta)$ does not depend on the input in cases where $s(x,z,\theta)=0$, and neither $L_{\hat f}h_x$ nor $L_{\hat g}h_x$ then introduces input dependence through $\phi$. Also, if the input is decomposed as $u=[u_x^T,u_z^T]^T$ with separate slow and fast actuation channels, the slow and fast adaptive CBF inequalities can likewise be enforced using $u_x$ and $u_z$, respectively, whenever the corresponding dynamics are channel-separated. These various cases discussed above, which are satisfied by many real-world systems, can simplify implementation of the adaptive CBF constraint.
\end{remark}

\section{Numerical Examples}
\label{sec:examples}
In this section, we will consider two numerical examples to illustrate
Theorems~\ref{thm:main} and \ref{thm:joint} and show both how the slow subsystem
safety can be robustly achieved using a CBF controller based on the
reduced-order dynamics and how joint slow-fast safety can be achieved using the
slow and fast subsystem CBFs simultaneously. Example~1 is a
synthetic fourth-order system while Example~2 considers a real-world example of a unicycle.

\subsection{Example 1: Synthetic fourth-order two-time-scale system with uncertain parameters}
Consider the uncertain system below with
$x=[x_1,x_2]^\top\in\mathbb{R}^2$ being the slow state,
$z=[z_1,z_2]^\top\in\mathbb{R}^2$ being the fast state, and
$u=[u_1,u_2]^\top$ being the control input vector with magnitude bounds
$|u_i|\le \overline U_i, i=1,2$:
\begin{align}
\dot{x}_1 &= \theta_1 x_2 + u_1 \ \ , \ \
\dot{x}_2 = x_1^2 z_2 + u_2 \nonumber\\
\varepsilon\dot{z}_1 &= -z_1 +u_1 \ \ , \ \
\varepsilon\dot{z}_2 = -z_2 + \theta_2 x_2^2 + u_2.
\label{eq:ex1}
\end{align}
Here $\theta=[\theta_1,\theta_2]^\top$ is an unknown constant parameter
vector assumed to be in the known box
$\Theta=[0.75,1.25]\times[0.75,1.25]$.
In this example system, the control enters both the slow and fast dynamics. Note that the coupling between
the slow and fast subsystems is nonlinear: the fast state enters the slow
dynamics via the $x_1^2z_2$ term, while the slow state enters the fast dynamics via
the $\theta_2x_2^2$ term. The slow manifold for a given parameter value is
\begin{equation}
z=\phi(x,u,\theta)= [u_1,\theta_2 x_2^2 + u_2]^\top.
\label{eq:ex1_manifold}
\end{equation}
In the adaptive controller, the reduced model and the corresponding CBF
quantities are evaluated using the online estimate $\hat\theta$.

\noindent{\em Safety constraints:} As safety constraints, we will require the slow and fast states to remain
\emph{outside} disks centered at given points
$c_x=[c_{x1},c_{x2}]^\top\in\mathbb{R}^2$ and
$c_z=[c_{z1},c_{z2}]^\top\in\mathbb{R}^2$ and with radii $R_x>0$ and $R_z>0$,
respectively. Hence, define slow and fast subsystem CBFs as (for $\nu= x,z$):
\begin{align}
 h_\nu(\nu)&=\|\nu-c_\nu\|^2 - R_\nu^2, & \mathcal{C}_\nu &= \{\nu\mid h_\nu(\nu)\ge0\}
 \label{eq:ex1_hx}
\end{align}
The action of these CBFs is that if the slow/fast states approach their
respective ``keep-out'' zones, then the CBF-based controller makes the state
move along gradients that point outward from these unsafe exclusion disks using control
actions that repel the state from entering the exclusion disks.

\noindent{\em Verification of assumptions:}
It can be verified that system \eqref{eq:ex1} satisfies all the Assumptions~\ref{asm:smooth}-\ref{asm:cbf_gradient_bound}.
Assumption~\ref{asm:smooth} is trivially satisfied.
Assumption~\ref{asm:phi_bounded} is satisfied since the slow manifold $\phi(x,u,\theta)=[u_1,\theta_2 x_2^2+u_2]^\top$ has the Jacobian $\partial\phi/\partial u$ as a constant 2x2 identity matrix and the Jacobians $\partial\phi/\partial x = \operatorname{diag}(0,2\theta_2 x_2)$ and $\partial\phi/\partial \theta = \operatorname{diag}(0,x_2^2)$, all of whose entries are uniformly bounded over the compact set $\mathcal{X}\times\mathcal{U}\times\Theta$.
Assumption~\ref{asm:faststab} is seen to be satisfied due to the stabilizing terms $-z_1$ and $-z_2$ in the fast dynamics.
Boundedness of trajectories follows from a stabilizing controller $u_{\rm nom}$ that will be defined below.
As discussed earlier, the practical application of the CBF-based controller is as a safety filter which modifies the control input from the nominal stabilizing controller when necessary to enforce the safety conditions.
Assumptions~\ref{asm:margin} and \ref{asm:fastcbf} on the slow and fast CBF conditions will be seen to be satisfied from the CBF analysis below.
Assumption~\ref{asm:mismatch} is satisfied since $\|x_1^2z_2-x_1^2(\theta_2 x_2^2+u_2)\|$ can be bounded as a class-$\mathcal{K}$ function of $\|z-\phi(x,u,\theta)\|=\sqrt{(z_1-u_1)^2+(z_2-\theta_2 x_2^2-u_2)^2}$, which trivially follows from boundedness of trajectories. In the adaptive controller, the reduced model is evaluated using the online estimate $\hat\theta$, and the resulting parameter mismatch is accounted for by the robustness margins $\sigma_x\beta_\theta$ and $\sigma_z\beta_\theta$ defined below. Assumption~\ref{asm:cbf_gradient_bound} is trivially satisfied since $h_z$ does not depend directly on $x$ for this example.

\noindent{\em Estimated reduced model and slow CBF derivative:}
With $z=\phi(x,u,\hat\theta)$, the estimated reduced-order dynamics for the slow subsystem are
\begin{align}
 \dot{x}_1 &= \hat\theta_1 x_2 + u_1 \ \ , \ \
 \dot{x}_2 = x_1^2\big(\hat\theta_2 x_2^2 + u_2\big) + u_2.
\end{align}
With $h_x(x)$ defined in \eqref{eq:ex1_hx}, we have $\nabla h_x=2(x-c_x)$ and
\begin{align}
 \dot h_x &= a_x(x,\hat\theta) + b_x(x)u
 \\
a_x(x,\hat\theta) &= 2(x_1-c_{x1})\hat\theta_1 x_2 +2(x_2-c_{x2})x_1^2\hat\theta_2 x_2^2
\\
b_x(x) &= [2(x_1-c_{x1}), 2(x_2-c_{x2})(1+x_1^2)]^\top.
\end{align}
The slow CBF condition is imposed as in Assumption~\ref{asm:margin} picking $\alpha_x(h)=\kappa_x h$ with $\kappa_x>0$, margin $\gamma_x>0$, and robustness margin $\sigma_x\beta_\theta>0$ as:
\begin{align}
a_x(x,\hat\theta)+b_x(x)u &\geq -\kappa_x h_x(x)+\gamma_x + \sigma_x\beta_\theta.
\label{eq:ex1_hxcond}
\end{align}
When only the slow CBF condition is considered, it is evident that inequality \eqref{eq:ex1_hxcond} is always feasible (i.e., a suitable $u$ can always be picked) by the fact that $\|b_x(x)\|$ is non-zero in the safe set $\mathcal{C}_x$ since $\|x-c_x\|\ge R_x$ in the safe set.

\noindent{\em Fast CBF derivative:} Considering ``frozen-$x$'' dynamics as in
Assumption~\ref{asm:fastcbf}, we have along the fast subsystem dynamics:
\begin{align}
 L_{r,z}h_z+L_{s,z}h_zu &=
 a_z(x,z,\hat\theta) + b_z(x,z)u
\end{align}
where
\begin{align}
  a_z(x,z,\hat\theta) &= 2 (z_1\!-\!c_{z1}) (-z_1) \!+\! 2 (z_2-c_{z2}) (-z_2 + \hat\theta_2 x_2^2) \\
  b_z(x,z) &= [2 (z_1-c_{z1}),2 (z_2-c_{z2})]^\top.
\end{align}
Picking
$\alpha_z$ to be the linear extended class-$\mathcal K$ function
$\alpha_z(h_z)=\kappa_z h_z$ with $\kappa_z>0$ and picking margin $\gamma_z>0$,
the fast CBF condition from Assumption~\ref{asm:fastcbf} (augmented with the
robustness margin $\sigma_z\beta_\theta>0$ to address the parameter mismatch in
the frozen slow state) is:
\begin{equation}
 a_z(x,z,\hat\theta) + b_z(x,z) u \ge -\kappa_z h_z(z) + \gamma_z + \sigma_z\beta_\theta.
\label{eq:fast_cbf_exact}
\end{equation}
Note that $\|x-c_x\|$ and $\|z-c_z\|$ are bounded below by $R_x$ and $R_z$, respectively, everywhere in $\mathcal{C}_{xz}$, implying that $b_x$ and $b_z$ are non-zero throughout $\mathcal{C}_{xz}$ (and linearly independent except on a set of measure zero in the fourth-order space, which can however be shown to not be aligned with the slow manifold, implying that any traversals through the set of measure zero will be instantaneous if they at all occur). Hence, the CBF inequalities \eqref{eq:ex1_hxcond} and \eqref{eq:fast_cbf_exact} can be satisfied simultaneously anywhere in $\mathcal{C}_{xz}$. Furthermore, since the safe set for the fast subsystem only requires keeping out of the unsafe circular region, it is seen that $u$ can be scaled appropriately if needed to ensure that $\phi(x,u,\hat\theta)$ remains outside the fast subsystem state exclusion disk, satisfying the last requirement in Assumption~\ref{asm:fastcbf}.

\noindent{\em Nominal controller, adaptive parameter update, and CBF filter using composite projection:}
As the nominal control law, we consider a simple stabilizing controller that ignores the exclusion disks and would (in the absence of CBF safety filters) drive the slow state toward the origin using the control policy given by
 $u_{\rm nom}(x) = [-k_1 x_1 , -k_2 x_2]^\top$.
The CBF conditions act as a safety filter to modify $u$ when necessary to keep the state outside the unsafe exclusion regions.
For this purpose, the actual control input $u$ is obtained by projecting $u_{\rm nom}$ as in Remark~\ref{rem:cbf_projection} as
 $u = \arg\min_{u\in\mathbb{R}^2} \ \|u-u_{\rm nom}\|^2$ under the inequality conditions in \eqref{eq:ex1_hxcond} and \eqref{eq:fast_cbf_exact} and verifying $h_z(\phi(x,u,\hat\theta))\geq 0$ at each time step as discussed in Remark~\ref{rem:cbf_projection}.

Since the unknown parameters enter the slow manifold affinely in $\theta_2$ and the slow dynamics affinely in $\theta_1$, the estimates $\hat\theta_1$ and $\hat\theta_2$ are updated online using a projected normalized-gradient law driven by the measured one-step finite-difference regressions of the corresponding uncertain terms. Specifically, at each time step $k$ during the simulation (with sampling time $\Delta_t$ set to $2\times 10^{-4}$ in the simulation), the parameter estimate is updated as
\begin{align}
 \hat\theta_1^{+} &= \Pi_\Theta\!\left(\hat\theta_1 + g_{\rm adapt}\Delta_t\,\frac{\phi_1\,e_1}{1+\phi_1^2}\right)
 \\
 e_1 &= \frac{x_{1,k+1}\!-\!x_{1,k}}{\Delta_t} - u_{1,k} - \hat\theta_1 \phi_1
 \\
 \hat\theta_2^{+} &= \Pi_\Theta\!\left(\hat\theta_2 + g_{\rm adapt}\Delta_t\,\frac{\phi_2\,e_2}{1+\phi_2^2}\right)
\\
 e_2 &= \varepsilon\,\frac{z_{2,k+1}\!-\!z_{2,k}}{\Delta_t} + z_{2,k} - u_{2,k} - \hat\theta_2 \phi_2
\label{eq:ex1_adapt}
\end{align}
with regressors $\phi_1=x_{2,k}$ and $\phi_2=x_{2,k}^2$, adaptation gain $g_{\rm adapt}>0$, and $\Pi_\Theta$ the projection onto the known parameter box $\Theta$. With this choice, $\hat\theta(t)\in\Theta$ for all $t\ge 0$ and the parameter-estimation error bound $\beta_\theta$ used in the margins can be taken as the worst-case deviation within the box, namely $\beta_\theta=\max_{\theta\in\Theta}\|\theta-\hat\theta\|=\sqrt{\sum_{i=1}^2[\max(\hat\theta_i-\theta_{lo,i},\theta_{hi,i}-\hat\theta_i)]^2}$, which is a known nonnegative function of the online estimate $\hat\theta$. Here, $\theta_{lo,i}$ and $\theta_{hi,i}$ are the known lower and upper bounds, respectively, for the two unknown parameters ($\theta_i, i=1,2$).

\begin{figure*}[t]
  \centering
  \includegraphics[width=\linewidth,trim=0.1in 0.1in 0.1in 0.1in, clip=true]{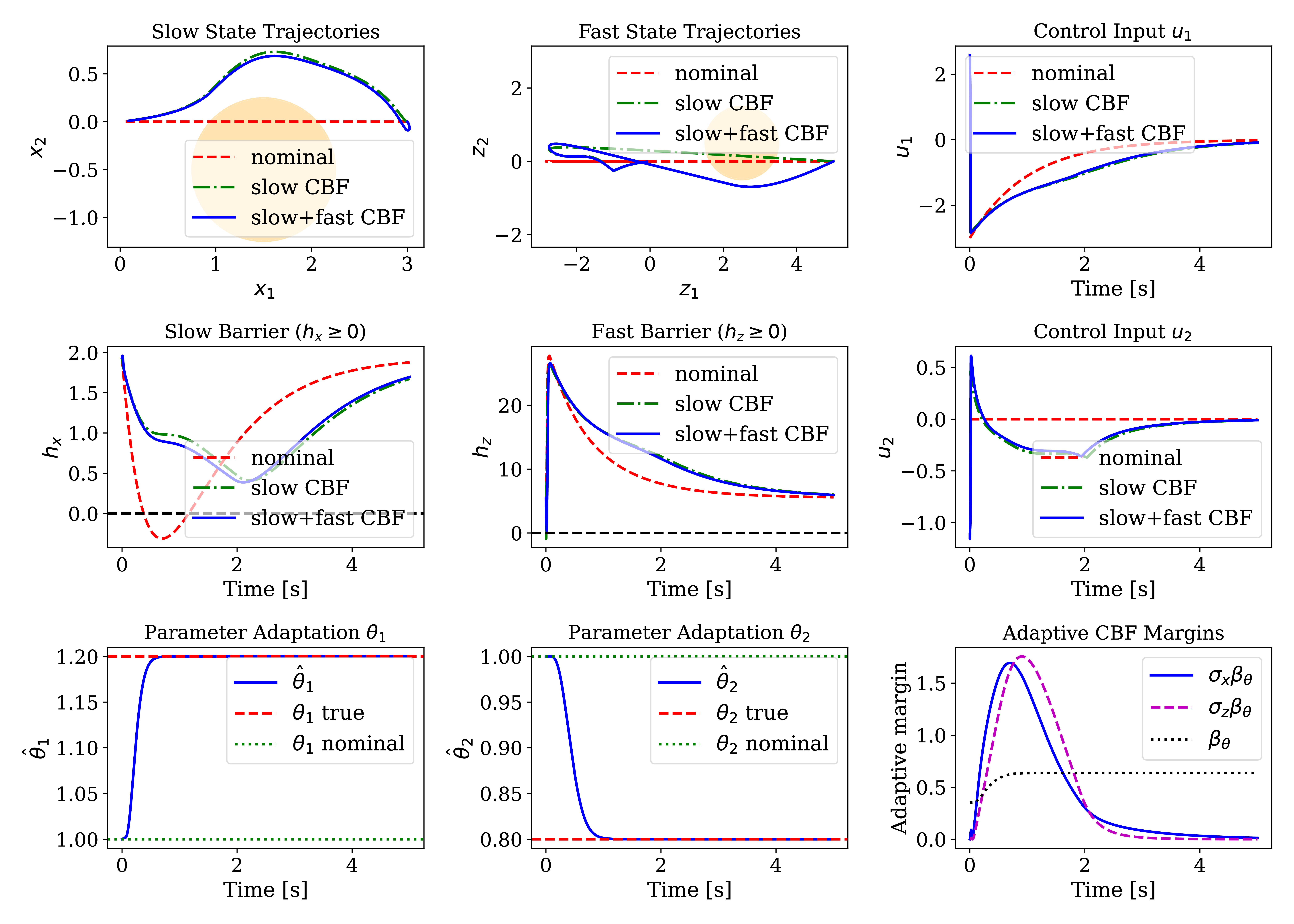}
  \caption{Simulation results for Example 1 under three cases: nominal controller (no CBFs) -- dashed red; slow CBF only -- dash-dotted green, both slow and fast CBFs -- solid blue. The plots show slow state trajectories in the $x$-plane with exclusion circle centered at $c_x$,
  fast state trajectories in the $z$-plane with exclusion circle centered at $c_z$,
  slow and fast CBFs $h_x$ and $h_z$, and control inputs $u_1$ and $u_2$. Note that in the slow state trajectory, CBF, and control input plots, the dash-dotted green lines overlap with the blue lines over most/all of the simulations.
  The bottom row additionally shows the online parameter adaptation: the estimates $\hat\theta_1$ and $\hat\theta_2$ (solid blue) together with their true values (dashed red) and nominal initial values (dotted green), and the adaptive CBF margins $\sigma_x\beta_\theta$ (solid blue), $\sigma_z\beta_\theta$ (dash-dotted magenta), and the parameter-estimation error bound $\beta_\theta$ (dotted black).
  }
  \label{fig:ex1}
\end{figure*}

\noindent{\em Simulation results:} For simulation studies, we use the parameters
 $c_x=[1.5,\ -0.5]^\top, R_x=0.75, c_z=[2.5,\ 0.5]^\top,R_z=1.0$ for the safe set definitions.
 The singular perturbation parameter $\varepsilon$ is set to 0.01.
 The unknown parameter vector is taken as $\theta=[1.2,\ 0.8]^\top$, which is assumed to lie in the known box $\Theta=[0.75,1.25]^2$ used by the projection $\Pi_\Theta$ and the adaptive margins. The parameter estimate is initialized at the nominal value $\hat\theta(0)=[1,\ 1]^\top$.
 For the controller parameters, we pick
 $k_1=k_2=1.0, \kappa_x=\kappa_z=2.0, \gamma_x=\gamma_z=0.01,\overline U_1=\overline U_2=6$, and the adaptation gain $g_{\rm adapt}=50$.
 We pick initial condition to be $x(0)=[3,0]^\top$ and $z(0)=[5,0]^\top$. Simulation results are shown in Figure~\ref{fig:ex1} for three cases: (i) only using the nominal controller defined above which attempts to drive $x$ to the origin; (ii) using only the slow CBF condition to filter the nominal controller's commands $u_{\rm nom}$; (iii) using both the slow and fast CBF conditions to filter $u_{\rm nom}$.
In cases (ii) and (iii), the reduced-order model and the CBF conditions are evaluated using the online estimate $\hat\theta$ updated by the projected normalized-gradient law \eqref{eq:ex1_adapt}, with the robustness margins $\sigma_x\beta_\theta$ and $\sigma_z\beta_\theta$ computed from the known parameter bound box.
The time series plots of these three cases are shown as dashed red, dash-dotted green, and solid blue, respectively, in Figure~\ref{fig:ex1}.
For each case, the plots show the slow state trajectories, fast state trajectories, control inputs $u_1$ and $u_2$, and the CBF values $h_x$ and $h_z$. In addition, the three adaptation-related panels in the bottom row show the evolution of the parameter estimates $\hat\theta_1$ and $\hat\theta_2$ (together with the true value in dashed red, and the nominal value in dotted green shown as references) and the adaptive CBF margins $\sigma_x\beta_\theta$ (solid blue), $\sigma_z\beta_\theta$ (dash-dotted magenta), and the parameter-estimation error bound $\beta_\theta$ (dotted black). The unsafe regions (exclusion circles) are shown in orange in the plots of the slow state and fast state trajectories.
 From the plots, a few observations can be seen. Firstly, when using the nominal controller, the slow state goes to the origin as desired, but the safety requirements are completely ignored. Secondly, when using only the slow CBF condition, the slow state safety requirement is satisfied (i.e., the slow state stays outside of its exclusion circle); but, the fast state safety requirement is violated briefly soon after time $t=0$. Thirdly, when using both slow and fast CBF conditions simultaneously, both slow and fast safety requirements are satisfied.
In addition, the parameter estimates $\hat\theta_1$ and $\hat\theta_2$ initialized at the nominal values $[1,1]^\top$ are seen to converge toward their true values $\theta_1=1.2$ and $\theta_2=0.8$. The adaptive margins $\sigma_x\beta_\theta$ and $\sigma_z\beta_\theta$ remain strictly positive throughout the transient, which preserves the robustness guarantees of the CBF conditions even before parameter convergence has been achieved.
These observations are as expected from the theoretical analysis and can be seen from both the state trajectories (in the $x$-plane and $z$-plane) and the CBF plots.
 The simulation results verify two properties as expected. Firstly, when a CBF controller is designed using the reduced-order slow subsystem dynamics where the fast state is replaced with its slow manifold approximation, the slow subsystem safety can be achieved even though the ignored fast dynamics is still present in the overall system. Secondly, when using slow and fast subsystem CBFs simultaneously, joint slow--fast safety is achieved even though the two CBFs both address only parts of the overall system.

\subsection{Example 2: Unicycle with fast turn-rate actuator (with uncertain parametric uncertainty) and including obstacle avoidance as a safety constraint}
Consider a unicycle dynamics with constant forward speed $v_0$ and a fast first-order actuator for the turn rate $\omega$. The actuator dynamics involve an unknown coefficient $\theta>0$ multiplying $\omega$, so the plant is given as
\begin{align}
\dot p_x = v_0 \cos\varphi,\
\dot p_y = v_0 \sin\varphi,\
\dot{\varphi} = \omega,\
\varepsilon\dot{\omega} = -\theta\,\omega + u
\end{align}
where $x=(p_x,p_y,\varphi)$ are the slow states and $z=\omega$ is the fast state. The parameter $\theta$ is unknown but is assumed to lie in a known positive interval $\Theta=[\theta_{lo},\theta_{hi}]\subset\mathbb{R}_{>0}$. For small $\varepsilon>0$, $\omega$ tracks the input divided by $\theta$ on the slow manifold given by $\omega=\phi(x,u,\theta)=u/\theta$, which is evaluated using the online estimate $\hat\theta$ in the adaptive controller.

\noindent{\em Safety constraints:}  We require the unicycle robot to avoid a circular obstacle with center at $[p_{xo},p_{yo}]^\top$ and radius $R_o$ while keeping the turn rate bounded (magnitude less than a positive constant $\omega_{\max}$). For this purpose, define the slow and fast subsystem CBFs as
\begin{align}
 h_{x}(x) &= (p_x-p_{xo})^2+(p_y-p_{yo})^2 - R_o^2\\
 h_{z}(z) &= \omega_{\max}^2 - \omega^2.
\end{align}

\noindent{\em ECBF construction and slow subsystem ECBF condition:} Unlike in Example 1, the slow subsystem CBF $h_x$ here has a relative degree of 2 with respect to the appearance of the control input via the slow manifold approximation, i.e., since $\dot\varphi=\omega\approx u/\hat\theta$ on the estimated slow manifold, $u$ appears only in the second derivative of $h_x$.
Hence, we will use an ECBF construction to enforce the slow subsystem safety. Define relative coordinates $p_{xr}=p_x-p_{xo}$, $p_{yr}=p_y-p_{yo}$, and
\begin{align}
 p &= p_{xr}\cos\varphi \!+\! p_{yr}\sin\varphi,\  q = -p_{xr}\sin\varphi + p_{yr}\cos\varphi.
\end{align}
Hence, we have
 $\dot h_x = 2v_0 p$ and
 $\ddot h_x = 2v_0 (v_0 + q\,\omega) = 2v_0(v_0 + q\,u/\hat\theta)$ on the estimated slow manifold.
An exponential CBF condition with gains $c_0,c_1>0$ and margin $\gamma_x>0$ (augmented with the robustness margin $\sigma_x\beta_\theta>0$ to address the parameter mismatch in the slow manifold) is given by
\begin{align}
\ddot h_x + c_1 \dot h_x  \ge -c_0 h_x + \gamma_x + \sigma_x\beta_\theta
\end{align}
which is an affine condition in $u$ of form $a + bu \ge 0$ with
\begin{align}
a &= 2v_0^2 + 2c_1v_0p + c_0 h_x + \gamma_x + \sigma_x\beta_\theta\\
b &= 2v_0 q / \hat\theta.
\end{align}
This ECBF condition can be used analogously to the slow CBF condition in Assumption~\ref{asm:margin}.

\noindent{\em Verification of assumptions:} Assumptions~\ref{asm:smooth} and Assumption~\ref{asm:faststab} are easily seen to be satisfied.
Assumption~\ref{asm:phi_bounded} is satisfied since the slow manifold $\phi(x,u,\theta)=u/\theta$ has the bounded Jacobians $\partial\phi/\partial u = 1/\theta\le 1/\theta_{lo}$ and $\partial\phi/\partial\theta = -u/\theta^2$, both of which are uniformly bounded on the compact set $\mathcal{X}\times\mathcal{U}\times\Theta$ (with $|u|\le \theta_{lo}\omega_{\max}$ enforced by the robust fast-manifold compatibility condition and $\theta\ge\theta_{lo}>0$).
Assumptions~\ref{asm:margin} and \ref{asm:mismatch} are seen to be satisfied with the ECBF modification as constructed above.
Analogously to Example 1, the upper bound in Assumption~\ref{asm:mismatch} follows from boundedness of trajectories, which is ensured by a nominal stabilizing controller $u_{\rm nom}$ defined below.
Assumption~\ref{asm:fastcbf} on the fast CBF condition will be seen to be satisfied from the CBF analysis below.
Assumption~\ref{asm:cbf_gradient_bound} is trivially satisfied since $h_z$ does not depend directly on $x$ for this example.

\noindent{\em Fast CBF derivative:} Analogous to Example 1, picking $\alpha_z(h_z)=\kappa_zh_z$, the adaptive fast CBF condition is obtained in the form \eqref{eq:fast_cbf_exact} with $a_z(x,z,\hat\theta) = 2\hat\theta\,\omega^2$ and $b_z(x,z) = -2\omega$, augmented with the robustness margin $\sigma_z\beta_\theta>0$ to address the parameter mismatch in the frozen fast dynamics.
Since the slow and fast CBF conditions are both affine in $u$, it can be seen that $u$ can be picked for any $(x,z)\in\mathcal{C}_{xz}$ to satisfy these conditions simultaneously (by taking intersections of the obtained lower-upper intervals).
The last condition of Assumption~\ref{asm:fastcbf} that $h_z(x,\phi(x,u,\hat\theta))\geq 0$ can be enforced by restricting $u$ to a maximum magnitude of $|u|\le \theta_{lo}\,\omega_{\max}$, which ensures $|u/\theta|\le\omega_{\max}$ for every $\theta\in\Theta$ on the slow manifold.

\noindent{\em Nominal controller, adaptive parameter update, and CBF filter using composite projection:}
The nominal controller is designed as a simple steering control law that attempts to drive the robot radially inward to the origin with a scaling based on distance to the origin. Defining $r = \sqrt{p_x^2+p_y^2}$ and $\psi = \operatorname{atan2}(p_y,p_x)$, the controller sets $\varphi_{\mathrm{des}} = \mathrm{wrap}(\psi+\pi)$ and
$u_{\mathrm{nom}} = k_{h0}\,\mathrm{wrap}(\varphi_{\mathrm{des}}-\varphi)\left(k_{h1} +
k_{h2}\,r\right)$ where
$\mathrm{wrap}$ is the function that wraps angles to $[-\pi,\pi)$ and $k_{h0},k_{h1},k_{h2}>0$ are controller gains. The robot velocity $v_0$ is constant until it reaches a small neighborhood of the origin at which point it is set to zero.
To apply the CBF safety filters, note that both the slow and fast CBF inequalities are of the form $a + b u \ge 0$, therefore defining half-spaces in $u$. Hence, intersecting the resulting intervals as well as the interval $[-\theta_{lo}\omega_{max},\theta_{lo}\omega_{max}]$ noted above for the turn rate constraint, the nominal heading control $u_{\mathrm{nom}}$ can be projected onto the feasible interval. This avoids solving a QP and gives exact solutions for scalar $u$.

The parameter estimate $\hat\theta$ is updated online using a projected normalized-gradient law driven by the measured finite-difference regression of the uncertain term $-\theta\omega$ in the fast dynamics. Specifically, at each time step $k$ during the simulation (with sampling time $\Delta_t=0.001$ in the simulations), the estimate is updated as
\begin{align}
 \hat\theta^{+} &= \Pi_\Theta\!\left(\hat\theta + g_{\rm adapt}\Delta_t\,\frac{\omega_k\,e_k}{1+\omega_k^2}\right)\\
 e_k &= u_k - \varepsilon\,\frac{\omega_{k+1}\!-\!\omega_k}{\Delta_t} - \hat\theta\,\omega_k,
\label{eq:ex2_adapt}
\end{align}
with adaptation gain $g_{\rm adapt}>0$ and $\Pi_\Theta$ the projection onto the known parameter box $\Theta=[\theta_{lo},\theta_{hi}]$. With this choice, $\hat\theta(t)\in\Theta$ for all $t\ge 0$ and the parameter-estimation error bound used in the margins can be taken as the worst-case deviation within the box, i.e., $\beta_\theta=\max(\hat\theta-\theta_{lo},\theta_{hi}-\hat\theta)$, which is a known nonnegative function of the online parameter estimate $\hat\theta$.

\begin{figure*}[t]
  \centering
  \includegraphics[width=\linewidth,trim=0.1in 0.1in 0.1in 0.1in, clip=true]{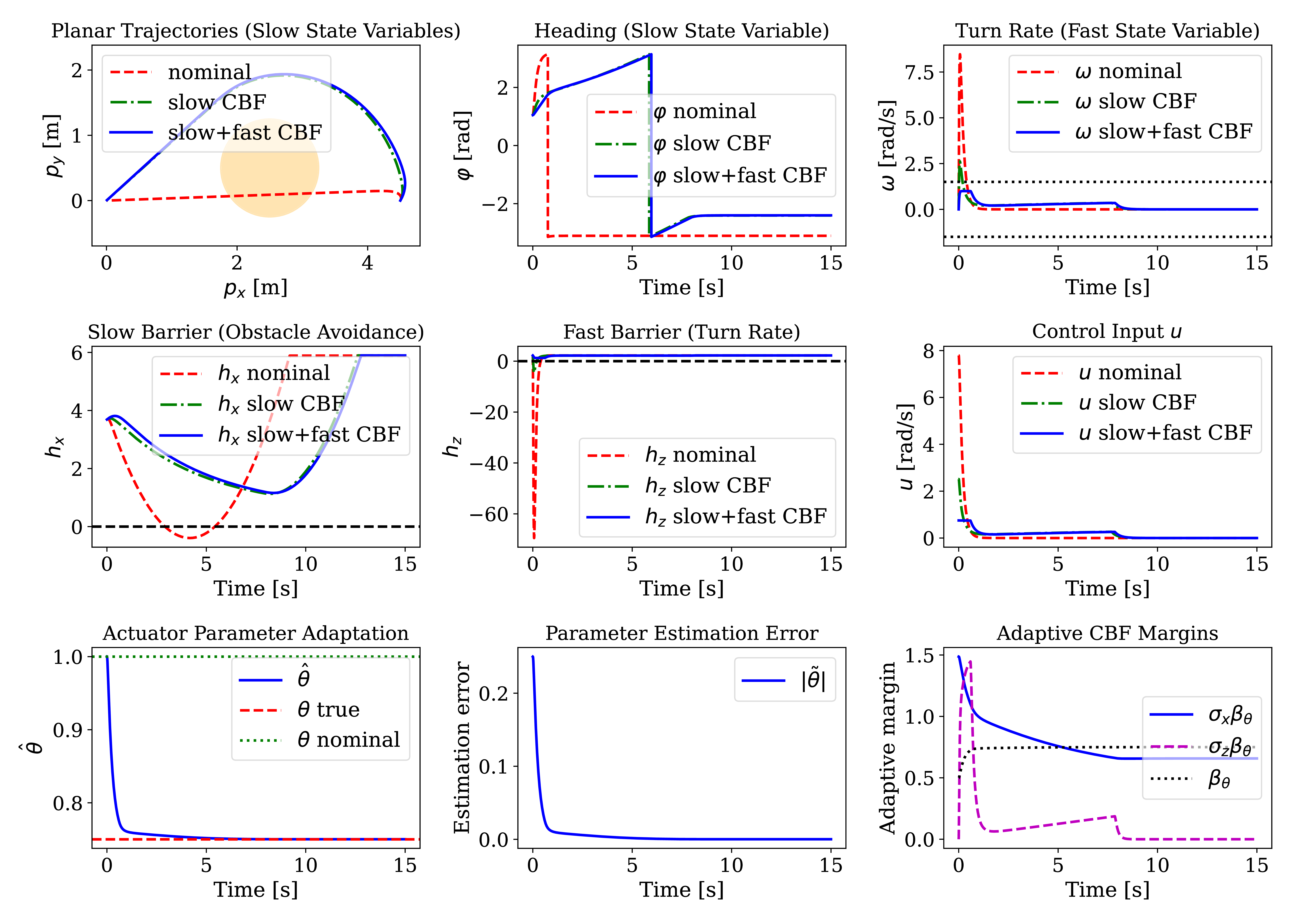}
  \caption{Simulation results for Example 2 under three cases: nominal controller (no CBFs) -- dashed red; slow CBF only -- dash-dotted green, both slow and fast CBFs -- solid blue. The plots show slow state planar trajectories ($p_x,p_y$) with obstacle (keep-out region shown as orange shaded circle), heading angle $\varphi$ (slow state variable),
  turn rate (fast state variable),
  obstacle avoidance CBF $h_x$ for slow subsystem, turn rate constraint CBF $h_z$ for fast subsystem, and control input $u$. Note that in slow state trajectories, CBF, and control input plots, dash-dotted green lines overlap with blue lines over most/all of the simulations.
  The bottom row additionally shows the online parameter adaptation: the estimate $\hat\theta$ (solid blue) together with the true value (dashed red) and the nominal initial value (dotted green), the parameter estimation error $|\hat\theta-\theta|$, and the adaptive CBF margins $\sigma_x\beta_\theta$ (solid blue), $\sigma_z\beta_\theta$ (dash-dotted magenta), and $\beta_\theta$ (dotted black).}
  \label{fig:ex3}
\end{figure*}

\noindent{\em Simulation results:}
For simulation studies, the parameters for the safety requirements are set to
$(p_{xo},p_{yo})=(2.5,\,0.5)$~m, $R_o=0.75$~m, $\omega_{\max}=1.5$~rad/s. The constant velocity is set to $v_0=0.5$~m/s.
The singular perturbation parameter $\varepsilon$ is set to 0.02.
The unknown actuator parameter is taken as $\theta=0.75$, and the known bounds for the unknown parameter are assumed to be given as $\Theta=[0.5,1.5]$ to be used by the projection $\Pi_\Theta$ and the adaptive margins. The parameter estimate is initialized at the nominal value $\hat\theta(0)=1.0$.
Controller parameters are set to
$k_{h0}=4.0$, $k_{h1}=0.3$, $k_{h2}=0.14$,
$\kappa_z=2.0$, $\gamma_x=0.01$, $\gamma_z=0.01$, $c_0=1.0$, $c_1=4.0$, and the adaptation gain $g_{\rm adapt}=8$.
Initial conditions are picked to be $p_x(0)=5.0$, $p_y(0)=0$, $\varphi(0)=\pi/3$, $\omega(0)=0$.
Simulation results are shown in Figure~\ref{fig:ex3} for three cases: (i) only using the nominal controller defined above which attempts to drive the robot to the origin (dashed red in Figure~\ref{fig:ex3}); (ii) using only the slow CBF condition to filter the nominal controller's commands $u_{\rm nom}$ to enforce obstacle avoidance (dash-dotted green in Figure~\ref{fig:ex3}); (iii) using both the slow and fast CBF conditions to filter $u_{\rm nom}$ to enforce the turn rate constraint (solid blue in Figure~\ref{fig:ex3}).
In all three cases, the projected normalized-gradient parameter update \eqref{eq:ex2_adapt} is executed online, so that cases (ii) and (iii) additionally act as adaptive CBF safety filters with the robustness margins $\sigma_x\beta_\theta$ and $\sigma_z\beta_\theta$ computed from the parameter box $\Theta$.
For each case, the plots show the slow and fast state trajectories, CBF values $h_x$ and $h_z$, and control input $u$. The obstacle is shown as shaded orange in the planar trajectories plot and the turn rate constraint is shown as black dotted lines in the turn rate plot. In addition, the three adaptation-related panels in the bottom row show the evolution of the parameter estimate $\hat\theta$ (with the true value shown as a dashed red line and the nominal initial value as a dotted green line), the parameter estimation error $|\hat\theta-\theta|$, and the adaptive CBF margins $\sigma_x\beta_\theta$ (solid blue), $\sigma_z\beta_\theta$ (dash-dotted magenta), and $\beta_\theta$ (dotted black).
From the plots, a few observations can be seen. Similar to Example 1, the nominal controller violates both the slow and fast safety requirements; the slow-only CBF case satisfies the obstacle avoidance requirement, but the turn rate constraint is violated, with $\omega$ exceeding the dotted $\omega_{\max}$ constraint because the slow-only case does not enforce the turn-rate CBF; and the joint slow+fast CBF case satisfies both obstacle avoidance and turn rate constraint safety requirements. In addition, the parameter estimate $\hat\theta$ initialized at the nominal value $1.0$ is seen to converge toward the true value $\theta=0.75$. The adaptive margins $\sigma_x\beta_\theta$ and $\sigma_z\beta_\theta$ remain strictly positive throughout the transient to preserve the robustness guarantees of the CBF conditions even before parameter convergence has been achieved. These observations are as expected and highlight two properties: the CBF controller designed using the reduced-order slow subsystem dynamics achieves slow subsystem safety even in the presence of the fast dynamics; joint slow--fast safety is achieved using slow and fast subsystem CBFs simultaneously.

\section{Conclusion}
\label{sec:conclusion}
An adaptive CBF-based framework was developed for nonlinear singularly perturbed systems with uncertain parameters
and two main results were shown: (i) preservation of slow subsystem safety properties under
a CBF controller designed using the reduced-order system even in the presence of
the fast dynamics; (ii) attaining of joint slow-fast safety using CBFs for slow
and fast subsystems simultaneously. Simulation studies with two example
systems were provided to illustrate the results, including a synthetic
fourth-order system and a unicycle with a fast turn-rate actuator and including obstacle avoidance as a safety constraint.
Topics for future work include analysis of wider classes of
nonlinear singularly perturbed systems including ones that are non-affine in
control inputs, joint design of stabilizing and CBF-based safety
controllers, tighter analysis of the bounds on the range of singular
perturbation parameters $\varepsilon^\star$ under which safety can be
guaranteed, robustness analysis to external disturbances, further analysis of conditions under which strict safety can be guaranteed, and experimental validation with real-world systems.

\section*{References}
\bstctlcite{IEEEexample:BSTcontrol}
\bibliographystyle{IEEEtran}
\bibliography{refs}
\bstctlcite{IEEEexample:BSTcontrol}

\end{document}